\documentclass[11pt,a4paper,nofootinbib]{article} 
\usepackage{jheppub}
\usepackage{enumerate}
\usepackage{epsfig}
\usepackage{amsmath} 
\usepackage{capt-of}
\usepackage{wasysym} 
\usepackage{mathrsfs} 
\usepackage{graphicx} 
\usepackage{amsfonts} 
\usepackage{amsbsy} 
\usepackage{amscd}
\usepackage{rotating}
\usepackage{slashed, latexsym, verbatim, cancel} 
\usepackage[utf8]{inputenc}
\usepackage{color}
\definecolor{myred}{rgb}{0.6,0,0} %usage:  {\textcolor{myred}{Hello World}}
\definecolor{myblue}{rgb}{0,0.2,0.4}
\definecolor{mygreen}{rgb}{0,0.9,0.1}
\definecolor{hc}{rgb}{.9,0.1,0.7}
\definecolor{hcout}{rgb}{.9,0.7,0.9}
\definecolor{Orange}{rgb}{1.,0.65,0.}
\makeatletter

\newcommand{\fmslash}[2][0mu]{%
  \mathchoice
    {\fmsl@sh\displaystyle{#1}{#2}}%
    {\fmsl@sh\textstyle{#1}{#2}}%
    {\fmsl@sh\scriptstyle{#1}{#2}}%
    {\fmsl@sh\scriptscriptstyle{#1}{#2}}}
\newcommand{\fmsl@sh}[3]{%
  \m@th\ooalign{$\hfil#1\mkern#2/\hfil$\crcr$#1#3$}}
\makeatother

\newcommand{\lsim}{{\;\raise0.3ex\hbox{$<$\kern-0.75em\raise-1.1ex\hbox{$\sim$}}\;}}
\newcommand{\gsim}{{\;\raise0.3ex\hbox{$>$\kern-0.75em\raise-1.1ex\hbox{$\sim$}}\;}}

\newcommand{\met}{{\fmslash E_T}}

\usepackage{array}
\newcolumntype{C}[1]{>{\centering\arraybackslash$}p{#1}<{$}}

\usepackage{bm} % bold in math mode
\newcommand{\be}{\begin{equation}}
\newcommand{\ee}{\end{equation}}
\newcommand{\bes}{\begin{equation*}}
\newcommand{\ees}{\end{equation*}}
\newcommand{\bea}{\begin{eqnarray}}
\newcommand{\eea}{\end{eqnarray}}
\newcommand{\beas}{\begin{eqnarray*}}
\newcommand{\eeas}{\end{eqnarray*}}

\newcommand{\wt}{\widetilde}
\def\ra{\rightarrow}
\title{Reconstructing a light pseudoscalar in the Type-X Two Higgs Doublet Model}
\author[a]{Eung Jin Chun,}
\author[b]{Siddharth Dwivedi,}
\author[b]{Tanmoy Mondal,} 
\author[b]{Biswarup Mukhopadhyaya} 
\affiliation[a]{Korea Institute for Advanced Study, Seoul 02455, Korea}
\affiliation[b]{Regional Centre for Accelerator-based Particle Physics,
Harish-Chandra Research Institute, HBNI,
Chhatnag Road, Jhunsi, Allahabad - 211 019, India} 
\emailAdd{ejchun@kias.re.kr}
\emailAdd{siddharthdwivedi@hri.res.in}
\emailAdd{tanmoymondal@hri.res.in}
\emailAdd{biswarup@hri.res.in}

\abstract{
We investigate the detectability as well as reconstructibility of a light pseudoscalar 
particle $A$, of mass in the 50 -- 60 GeV range, which is still allowed in a Type-X (lepton-specific)
two-Higgs doublet scenario. Such a pseudoscalar can be pair-produced in the 
decay $h\to AA$ of the 125 GeV scalar $h$. The light pseudoscalar in the aforementioned 
range, helpful in explaining the muon anomalous magnetic moment, has not only substantial
branching ratio in the $\tau^+\tau^-$ channel but also one of about $0.35\%$ in the 
$\mu^+\mu^-$ final state. We show how to faithfully reconstruct the $A$ mass using  the $\mu^+\mu^-$
mode, and establish the existence of a pseudoscalar around 50 -- 60 GeV, using 
the process $pp\to h \to AA\to \mu^+\mu^-\,\tau^+\tau^-$. This is the most reliable 
way of reconstructing the light $A$ mass, with a statistical significance that amounts 
to discovery, with a few hundred (or less) fb$^{-1}$ of integrated luminosity.}

\preprint{HRI-RECAPP-2017-009\\$\textrm{}$\hfill \today}


\begin{document}
\maketitle

\newpage

\section{Introduction}

Extension of the electroweak symmetric sector of the standard model
(SM) to two or more Higgs doublets is a widespread curiosity, of which
two Higgs doublet models (2HDM) occupy the centre stage. Such models in
general suffer from the flavour changing neutral current (FCNC)
problem. A popular way of avoiding FCNC is to use some discrete
symmetry (or something that effectively leads to it), which restricts
the Yukawa interactions of the two doublets. Based on the nature of
such symmetry, four types of 2HDM are popular, namely, Type-I,
Type-II, Type-X (or lepton specific) and Type-Y (or flipped)
\cite{Gunion:1989we,Djouadi:2005gj,Branco:2011iw}. This
paper contains some observations related to Type-X 2HDM.

In this scenario, one scalar doublet in the flavour basis has Yukawa
couplings with quarks only, while the other one couples to leptons
alone (Yukawa coupling with neutrinos are neglected without affecting
other aspects of phenomenology). The physical states other than the
SM-like 125-GeV scalar, obtained on diagonalizing the mass matrices,
have very small coupling with quarks compared to those with leptons,
once all constraints including those from the Large Hadron Collider
(LHC) are taken into account. This considerably relaxes the lower
bounds on some of the physical masses. In particular, it has been
found \cite{Broggio:2014mna,Chun:2015hsa,Chun:2016hzs} that the 
neutral pseudoscalar $A$ in Type-X 2HDM can be as
light at 40-60 GeV or even lighter in certain regions in the
parameter space, thanks to its generally low direct production rate at
the LHC and other colliders that have run so far. 
\footnote{Such light pseudoscalars may also occur in further extensions of the 
SM \cite{Bandyopadhyay:2015oga,Bandyopadhyay:2015tva, Goncalves:2016qhh}.}
And it is in part of these regions where the one-loop contribution induced by a light $A$
helps a good fit of the muon anomalous magnetic moment, especially for
high ($\ge 40$) values of $\tan\beta$, the ratio of the vacuum
expectation values of the two doublets~
\cite{Cheung:2001hz,Cheung:2003pw,Jegerlehner:2009ry,Lisanti:2009uy,Cao:2009as}.  
It is therefore important not
only to look for LHC signals of this scenario \cite{Chun:2015hsa}, but also to {\em
actually reconstruct the mass of the light A}. We suggest a method of
doing precisely that.

The light pseudoscalar, for large $\tan\beta$, has a $\tau^+ \tau^-$
branching ratio close to unity, and a $\mu^+ \mu^-$ branching ratio on
the order of $0.35 \%$. Signals have been suggested in the multi-tau channels like  $pp
\to HA \to \tau^+ \tau^- ~
\tau^+ \tau^-$ \cite{Su:2009fz,Kanemura:2011kx,Kanemura:2014bqa,Chun:2015hsa}. 
However, the taus cannot be reconstructed in the
collinear approximation \cite{Rainwater:1998kj} since there are four neutrinos in the
final state. Besides, even if only one $A$ decays into a $\tau$-pair,
the visible $\tau$-decay product (like a $\tau$-induced jet) cannot be
treated in the collinear approximation at such low energies as that
possessed by the $\tau$ produced from an $A$ as light as 50 -- 60
GeV. Therefore, we cannot reliably obtain $m_A$ using the
$\tau$-pair(s).  We find that the $\mu^+ \mu^-$ pair can come to one's
rescue here. With $pp \to hX \to AA
\to \tau^+ \tau^- ~ \mu^+ \mu^-$, one may reconstruct
$m_A$ from the muon pair, in association with a pair of tau-jets. We
show after a detailed simulation that such a strategy, combined
with that for suppressing SM backgrounds, isolates the signal events
carrying clear information on the pseudoscalar mass.  It is thus 
possible to achieve discovery-level statistical significance with an
integrated luminosity of about 100 $\rm{fb}^{-1}$ or less at 14 TeV.

In Section 2 we discuss the generic features of the
Type-X 2HDM with respect to the structure of the Yukawa and gauge couplings
of the physical scalars and point out how the parameter space of the model gets
constrained by the muon $g-2$ and precision observables. Section 3 is devoted 
to the the LHC analysis of our signal that identifies the pseudoscalar resonance, 
detailing the event selection criteria that helps in suppressing the backgrounds. 
Section 4 includes a discussion of the results
in the context of the efficacy of the analysis scheme used for our signal. 
We summarize and conclude in Section 5.

\section{The Type-X 2HDM Model and Constraints}
The  Type-X 2HDM with $\Phi_{1,2}$ as the two doublets is characterised by the following
Yukawa structure:
\begin{equation}\label{eq:yukawa}
{\cal L}_Y= -Y^u\bar{ Q_L} \wt \Phi_2 u_R + Y^d  \bar{ Q_L} \Phi_2 d_R+Y^e\bar{ l_L} \Phi_1 e_R + h.c.,
\end{equation}
where family indices are suppressed and  $\wt \Phi_2=i\sigma_2\Phi_2^*$. This Yukawa 
Lagrangian is the result of a $\mathbb{Z}_2$ symmetry~\cite{PhysRevD.15.1958}
which prevents tree level flavor changing neutral current. 
Under $\mathbb{Z}_2$,  the fields transform as $\Phi_2\ra \Phi_2$ and $\Phi_1\ra-\Phi_1$ combined with $e_R\ra -e_R$ while 
the other fermions are even under it.
Thus  $\Phi_2$ couples only to the quarks whereas 
$\Phi_1$ couples exclusively to the leptons.
The most general form of the  scalar potential is 
\begin{eqnarray}
\nonumber V_{\mathrm{2HDM}} &=& m_{11}^2\Phi_1^{\dagger}\Phi_1 + m_{22}^2\Phi_2^{\dagger}\Phi_2 -\Big[m_{12}^2\Phi_1^{\dagger}\Phi_2 + \mathrm{h.c.}\Big]
+\frac{1}{2}\lambda_1\left(\Phi_1^\dagger\Phi_1\right)^2+\frac{1}{2}\lambda_2\left(\Phi_2^\dagger\Phi_2\right)^2 \\
\nonumber && +\lambda_3\left(\Phi_1^\dagger\Phi_1\right)\left(\Phi_2^\dagger\Phi_2\right)+\lambda_4\left(\Phi_1^\dagger\Phi_2\right)\left(\Phi_2^\dagger\Phi_1\right)
+\Big\{ \frac{1}{2}\lambda_5\left(\Phi_1^\dagger\Phi_2\right)^2+\Big[\lambda_6\left(\Phi_1^\dagger\Phi_1\right) \\
&& +\lambda_7\left(\Phi_2^\dagger\Phi_2\right)\Big]\left(\Phi_1^\dagger\Phi_2\right) + \rm{h.c.}\Big\},
\label{eq:2hdmgen}
\end{eqnarray}
where all the couplings are assumed to be real. 
The $\mathbb{Z}_2$ symmetry implies $\lambda_6=\lambda_7=0$. However, the  term proportional to $m_{12}^2$,
which  softly breaks $\mathbb{Z}_2$ can be non zero to keep the quartic coupling $\lambda_1$ 
below perturbativity limit \cite{ Gunion:1989we,Gunion:2002zf}.   
%  since we are considering a CP-conserving 2HDM scenario. 
Parameterizing the doublets as $\Phi_j=(\phi_j^+,(v_j+\phi^r_j+i\phi^i_j)/\sqrt{2})^T$, we obtain 
the five physical massive states $A$, $h$, $H$, $H^{\pm}$ in terms of the two diagonalizing angles 
$\alpha$ and $\beta$:
\begin{eqnarray}\label{}
A&=&-s_\beta \;\phi^i_1+c_\beta \;\phi^i_2,\quad H^+=-s_\beta\; \phi_1^+ +c_\beta\; \phi^+_2,\cr
h&=&-s_\alpha\; \phi^r_1+c_\alpha\; \phi^r_2,\quad~ H=c_\alpha\; \phi^r_1+s_\alpha\; \phi^r_2,
\end{eqnarray}
where $s_\alpha$ and $c_\beta$ stand for $\sin\alpha$ and $\cos\beta$, etc. The CP-even state 
$h$ corresponds to the SM-like Higgs with mass $M_h=125$ GeV. Furthermore we look for the mass
hierarchy $M_A < M_h < M_H\simeq M_{H^\pm}$ which can be realised by setting $\lambda_4 + \lambda_5 \approx 0$. 
The SM-like Higgs couples to the pseudoscalar with strength
$\lambda_{hAA} = -(\lambda_3 + \lambda_4 - \lambda_5)v$, where  $v = \sqrt{{v_1}^2 + {v_2}^2} = 246$ GeV.

The Yukawa Lagrangian of Eq.(\ref{eq:yukawa}) can be rewritten in terms of the physical Higgs bosons, 
$h, H, A$ and $H^\pm$:
\begin{eqnarray}
\nonumber \mathcal L_{\mathrm{Yukawa}}^{\mathrm{Physical}} &=&
-\sum_{f=u,d,\ell} \frac{m_f}{v}\left(\xi_h^f\overline{f}hf +
\xi_H^f\overline{f}Hf - i\xi_A^f\overline{f}\gamma_5Af \right) \\
 &&-\left\{ \frac{\sqrt{2}V_{ud}}
{v}\overline{u}\left(m_{u}\xi_A^{u}P_L+m_{d}\xi_A^{d}P_R\right)H^{+}d  +
\frac{\sqrt{2}m_l}{v}\xi_A^l\overline{v}_LH^{+}l_R + \mathrm{h.c.}\right\},
\label{eq:L2hdm}
\end{eqnarray}
where $f$ runs over all of the quarks and charged leptons, and  $u$, $d$, and $l$ refer to the 
up-type quarks, down-type quarks, and charged leptons, respectively. 
The multiplicative factors of the 
Yukawa couplings, {\it i.e.} $\xi_h^f$, $\xi_H^f$ and $\xi_A^f$ are given in Table \ref{Tab:YukawaFactors}. For $\sin(\beta-\alpha) \approx 1 $ the Yukawa coupling with the SM-like 
Higgs $(h)$ are similar to that of the SM. 
\begin{table}[t]
%-------------------------------------------------------------------------------
\begin{center}
\begin{tabular}{|c||c|c|c|c|c|c|c|c|c|}
\hline
& $\xi_h^u$ & $\xi_h^d$ & $\xi_h^\ell$
& $\xi_H^u$ & $\xi_H^d$ & $\xi_H^\ell$
& $\xi_A^u$ & $\xi_A^d$ & $\xi_A^\ell$ \\ \hline
Type-X
& $c_\alpha/s_\beta$ & $c_\alpha/s_\beta$ & $-s_\alpha/c_\beta$
& $s_\alpha/s_\beta$ & $s_\alpha/s_\beta$ & $c_\alpha/c_\beta$
& $\cot\beta$ & $-\cot\beta$ & $\tan\beta$ \\
 \hline
\end{tabular}
\end{center}
 \caption{The multiplicative factors of Yukawa interactions in type X 2HDM}
\label{Tab:YukawaFactors}
\end{table}
In any type of the 2HDM, the  couplings of scalars with a pair of gauge bosons are given by
\cite{Gunion:1989we,Djouadi:2005gj, Kanemura:2014dea}: 
\begin{equation}
g_{hVV}=\mathrm{sin}(\beta-\alpha)g_{hVV}^{\mathrm{SM}},\,\,\,\,g_{HVV}=\mathrm{cos}(\beta-\alpha)g_{hVV}^{\mathrm{SM}},\,\,\,\,g_{AVV}=0,
\end{equation}
where $V$ = $Z,\,W^\pm$. 
The couplings of $Z$ boson with the neutral scalars are,
\begin{align}
&hAZ_\mu:\,\frac{g_Z^{}}{2}\cos(\beta-\alpha)(p+p')_\mu,\quad
 HAZ_\mu:\,-\frac{g_Z^{}}{2}\sin(\beta-\alpha)(p+p')_\mu,
\label{hhV}
\end{align}
where $p_\mu$ and $p'_\mu$ are
outgoing four-momenta of the first and the second scalars, respectively,
and $g_Z^{}=g_W^{}/\cos\theta_W^{}$.

For reasons already stated, we are concerned with the region corresponding to
$ \tan{\beta}\equiv v_2/v_1\gg 1$. This is because the  contribution
to the muon {$g-2$} coming from the Barr-Zee \cite{PhysRevLett.65.2626} two-loop diagrams 
can be substantial with a light pseudoscalar $A$ and 
$\tau$ running in the loops. Constraints on 2HDM parameter space coming from $(g-2)_\mu$ 
have been analyzed in 
Refs.~\cite{Dedes:2001nx,Cheung:2001hz,Krawczyk:2001pe,Krawczyk:2002df,Cheung:2003pw,Broggio:2014mna,Wang:2014sda,Chun:2015hsa,Ilisie:2015tra,Abe:2015oca,Han:2015yys,Cherchiglia:2016eui}
 and it was shown 
 in the updated analysis ~\cite{Chun:2016hzs}
that light $A$ in Type-X 2HDM can explain $(g-2)_\mu$ at 2$\sigma$ while evading collider as well as
precision data constraints. 
While it is true that in the Type-II 2HDM a light pseudoscalar can explain the $(g-2)_\mu$ anomaly, 
there the lower bound on the charged Higgs mass is  $M_{H^+}>580$ GeV 
coming from the $B\to X_s\gamma$ measurement~\cite{Belle:2016ufb}. Such a  heavy charged Higgs 
is not compatible with the requirement of a light pseudoscalar~\cite{Broggio:2014mna}. 
Similarly, in Type-I and Type-Y 2HDM, too, a very light pseudoscalar and its enhanced 
coupling with the muons are not consistent, since that would also imply comparably 
strong coupling to at least  one type of quarks, leading to unacceptably large 
$A$ production at hadron colliders. 
Beside those models where the $A$ couples to muons proportionally to $\cot\beta$ cannot
explain $(g-2)_{\mu}$, since $\tan\beta \leq 1$ is disfavoured by a number of considerations.
It is only in the Type-X that a light $A$ 
can have enhanced coupling to the $\mu$ and the $\tau$, concomitantly suppressed 
coupling to all quarks, and all  phenomenological and other theoretical constraints
  (vacuum stability, perturbativity etc.) duly satisfied \cite{Staub:2017ktc}. Keeping this in mind,
we proceed to find a strategy for reconstructing $M_A$ at the LHC.

\section{ Signal of a light $A$ : An analysis for the LHC}
The light pseudoscalar in Type-X 2HDM can be produced at the LHC via associated production along with
the SM Higgs and also via the decay of the SM like Higgs. The associated production is proportional to 
$\cos^2(\beta-\alpha)$ and is suppressed  for $(\beta-\alpha) \simeq \pi/2$,
leaving $h\to A\,A$ as the dominant production mode for the pseudoscalar. 
The pseudoscalar  is lepto-philic and almost exclusively 
decays to $\tau$ lepton for large $\tan\beta$ with 
a very small branching ratio  to di-muon ($BR(A\to \mu\mu)\simeq (m_\mu/m_\tau)^2\simeq 0.35\%$). 
This will lead to copious production of four-$\tau$ events ($AA \to \tau^+\tau^-\;\tau^+\tau^-$), 
the characteristic Type-X signal which was analyzed in detail in 
Refs.~\cite{Su:2009fz,Kanemura:2011kx,Kanemura:2014bqa,Chun:2015hsa}. 
Since the decay of the $\tau$ involves neutrinos, full reconstruction of the four-$\tau$ system is 
not possible which rules out any possibility of identifying a resonance peak. On the other hand if we
consider the decay $AA\to \mu^+\mu^-\;\tau^+\tau^-$ it is straightforward to identify the events owing to 
clean di-moun invariant mass ($M_{\mu\mu}$) peak at $M_A$ which will be the `smoking gun' signal for a 
light spin-0 resonance. 
We show later that, in spite of the limited branching ratio for $A \to \mu^+\mu^-$, the $2\mu\;2\tau$ final 
state can identify the $A$ peak well within the luminosity reach of the 14 TeV LHC.

% \subsection{Signal}
\begin{table}[t]
\begin{center}
\begin{tabular}{|c|c|c|c|c|}\hline
  Parameters &  $M_{A}$ (GeV) & $\tan\beta$ & $\cos(\beta-\alpha)$  & $\lambda_{hAA}/v$ \\\hline\hline
  BP1        &  50	      &	60	    & 0.03  	 	   & 0.02 		  \\\hline
  BP2        &  60            & 60	    & 0.03		   & 0.03		  \\\hline
\end{tabular}
\end{center}
\caption{Benchmark points for studying the discovery prospects of light pseudoscalar 
  in Type X 2HDM model at 14 TeV run of LHC. $\lambda_{hAA}$ is in units of $v = 246$ GeV.}
\label{tab:benchmark}
\end{table}

The signal we are exploring  contains a pair of oppositely charged muons with exactly two 
$\tau$--tagged jets produced via :
\begin{equation}
  p\,p\to h\to A\,A\to \mu^+\mu^-\;\;\tau^+\tau^- \to \mu^+\mu^-\;\; {j_\tau}\,{j_\tau} +\met,
\end{equation}
where $j_\tau$ is a $\tau$--tagged jet as a result of hadronic $\tau$-decay. 
 The NNLO cross section for the Higgs production via gluon fusion at 14 
TeV LHC is $ 50.35$ pb \cite{higgs:xsection}. 

The Type-X 2HDM model  have been encoded using \texttt{FeynRules}~\cite{Christensen:2008py,Alloul:2013bka} 
in order to generate the model files for implementation in $\texttt{MadGraph5\_aMC@NLO}$~\cite{Alwall:2011uj,Alwall:2014hca} 
which was used for computing the required cross-sections and generating events for collider analyses.

 We have chosen the benchmark points (BP) given in Table \ref{tab:benchmark} for our analysis.
 As we have explained in the previous section, we want a light pseudoscalar which can explain the 
 muon $g-2$ anomaly at 2$\sigma$. 
The benchmark points in the parameter space used here, corresponding to $M_A = 50, 60$ GeV, are 
consistent with all phenomenological constraints. They also satisfy theoretical constraints 
such as perturbativity and a stable electroweak vacuum \cite{Broggio:2014mna}. 
The signal of a light $A$, which is our main focus here, does not depend on $M_H$ or $M_{H^\pm} $.
For both of our benchmark points, each of these masses is 200 GeV.  For the chosen  benchmark scenarios, 
the branching ratio of Higgs to $AA$ is $BR(h\to A A)\simeq 15\%$ which is well below  the upper limit of about 
23$\%$ \cite{Aad:2015pla} on any non-standard decay branching ratio (BR) of the  SM-like Higgs boson. 
The choice of $\tan\beta$ ensures that the lepton universality bounds originating from 
 $Z$ and $\tau$ decays are satisfied \cite{Chun:2016hzs}.

\subsection{Backgrounds}
The major backgrounds to our signal process : $ \mu^+\mu^-\,j_\tau \,j_\tau$ come from the following channels
(A) $p p \to \mu^+ \mu^- + jets$, (B) $p p \to V V + jets (V = Z,W, \gamma^*) $ and (C) $p p \to t \bar t + jets$.
All the background events are generated with two additional partons and the events are matched up to 
two jets using MLM matching scheme \cite{Mangano:2006rw,Hoche:2006ph}  using the \emph{shower-kT} algorithm 
with $p_T$ ordered showers.  We use NNLO production cross section for $\mu^{\pm}\,\mu^{\mp}\,j\,j$
\cite{Catani:2009sm} and $ZZ$ \cite{Cascioli:2014yka}, whereas  $t\,\bar t$ production cross section 
is computed at  N$^3$LO \cite{Muselli:2015kba}.
%----------------------------------------------------------------------------------------------------

Apart from these three backgrounds there exist other SM processes like  $VVV$,
$t\bar{t}V$ and $W^\pm Z$ which in principle could fake the proposed signal ($2\mu2\tau$)~\cite{CMS-PAS-HIG-16-036}.
However lower cross-section and the requirement of exactly two muons and two tau-tagged jets satisfying a tight 
invariant mass window around the pseudoscalar mass effectively eliminates the contribution from these additional channels.

\subsection{Simulation and event selection}\label{sec:simulation}

 \begin{figure}[t!]
 \includegraphics[width = 7.5cm]{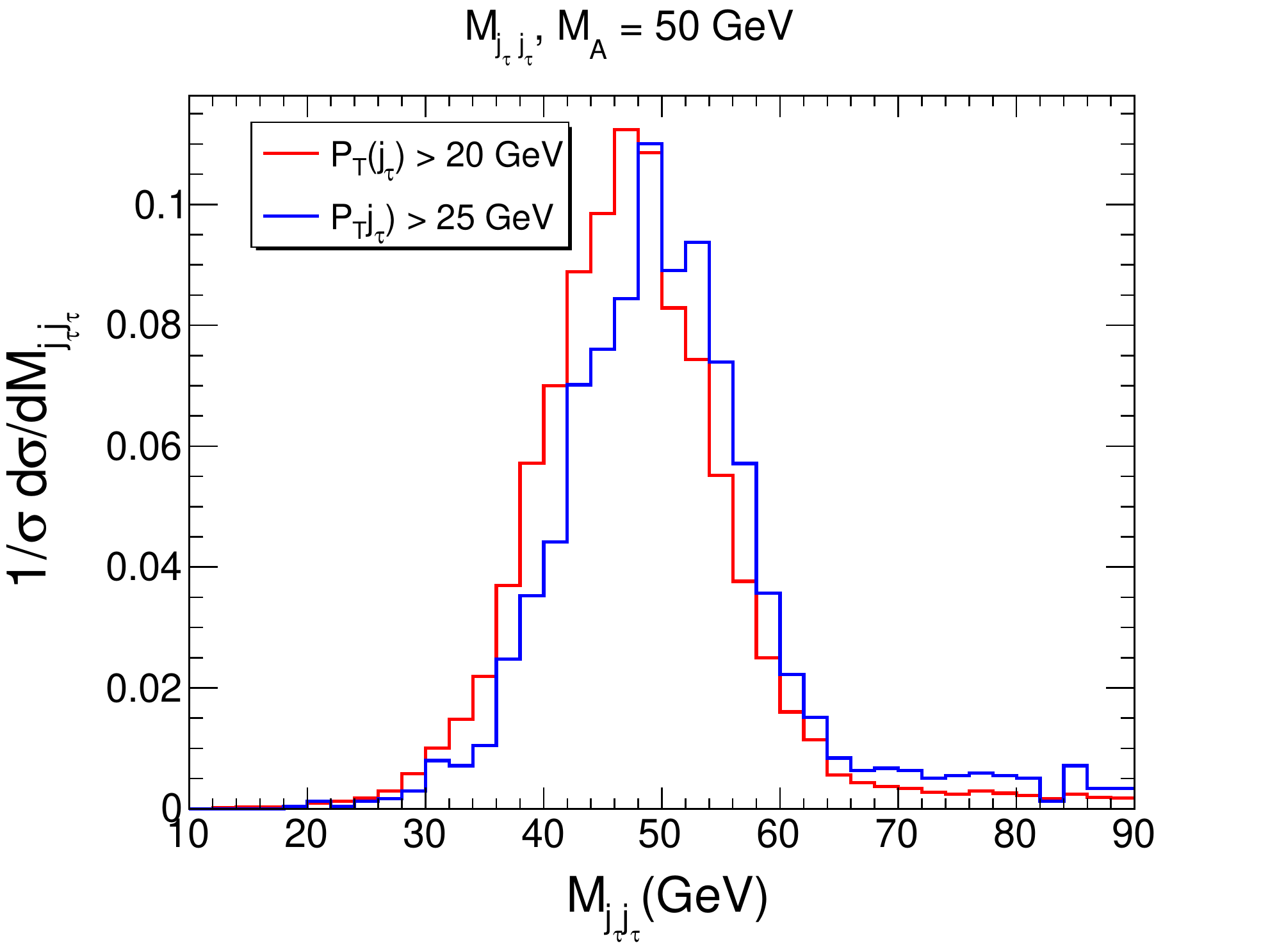}
  \includegraphics[width = 7.5cm]{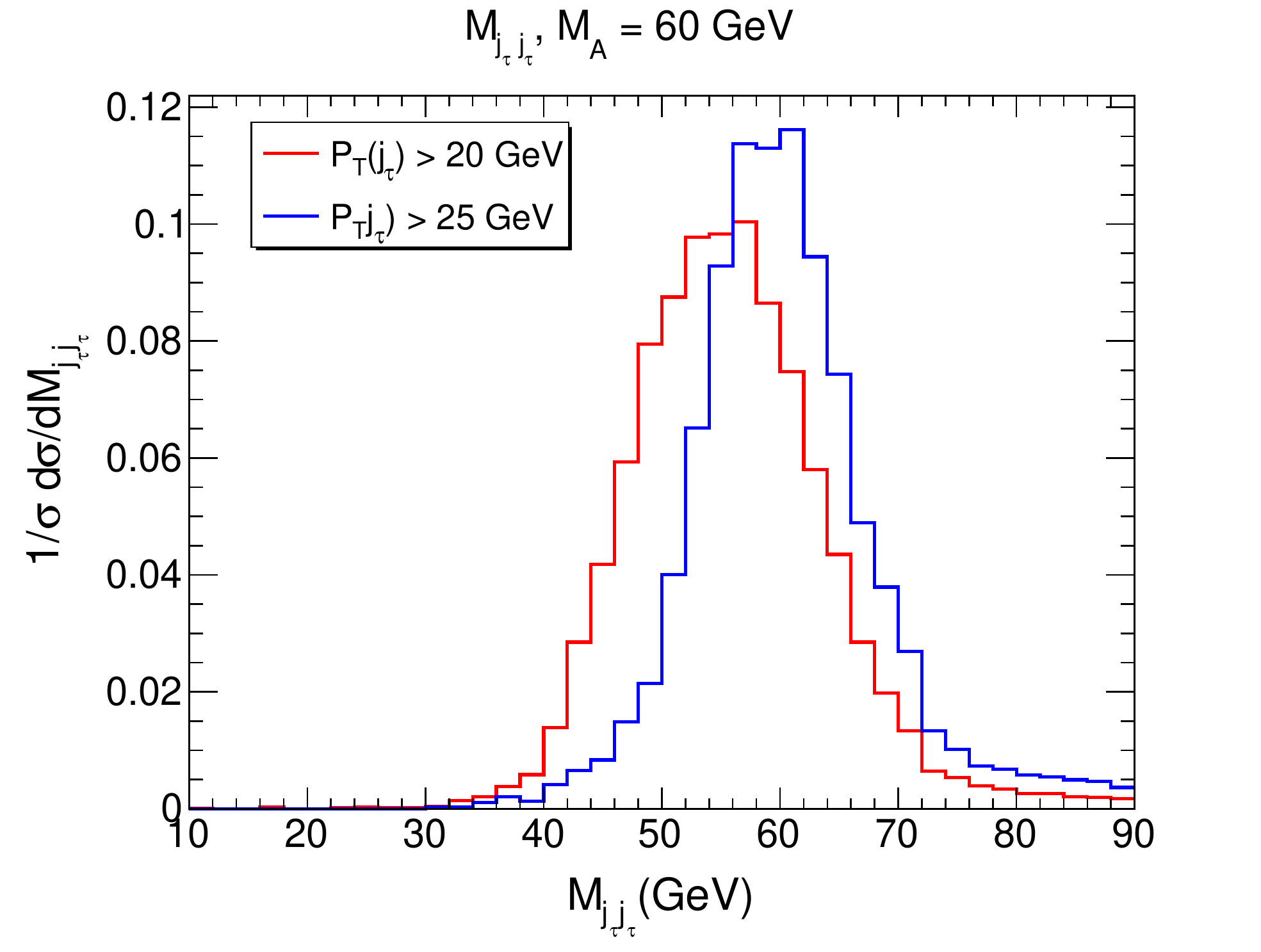}
   \caption{The invariant mass of the 2 tau-tagged jets for $M_A = 50$ and $M_A = 60$.
   The figures illustrate how the higher $p_T (j_\tau)$ threshold leads to more
   precise reconstruction of the peak at $M_A$.}
  \label{fig:IM_taujets}
  \end{figure}

After generating both signal and background  events with $\texttt{MadGraph5\_aMC@NLO}$, we have used 
\texttt{PYTHIA6} \cite{Sjostrand:2006za} for the subsequent decay, showering and hadronization 
of the parton level events. Decay of $\tau$ leptons is incorporated using \texttt{TAUOLA} \cite{Jadach:1993hs}
integrated in $\texttt{MadGraph5\_aMC@NLO}$. Both one- and three-prong   $\tau$ decays have been 
included in our analysis.
For  event generation we have used the \texttt{NN23LO1} \cite{Ball:2014uwa} parton distribution 
function and the default dynamic renormalisation and factorisation scales \cite{mad:scale} in $\texttt{MadGraph5\_aMC@NLO}$. 
Finally, detector simulation was done using \texttt{Delphes3} \cite{deFavereau:2013fsa}. 
Jets were reconstructed using the  anti-kT algorithm \cite{Cacciari:2008gp} with $R = 0.4$. The $\tau$-tagging
efficiency and mistagging efficiencies of the light jets as $\tau$-jets are incorporated in \texttt{Delphes3} 
as reported by the ATLAS collaboration~\cite{ATL-PHYS-PUB-2015-045}. We operate our simulation on 
the Medium tag point for which the tagging efficiency of 1-prong (3-prong) $\tau$ decay is 70\% (60\%)
and the corresponding mistagging rate is 1\% (2\%).

\begin{figure}[t]
 \includegraphics[width = 7.5cm]{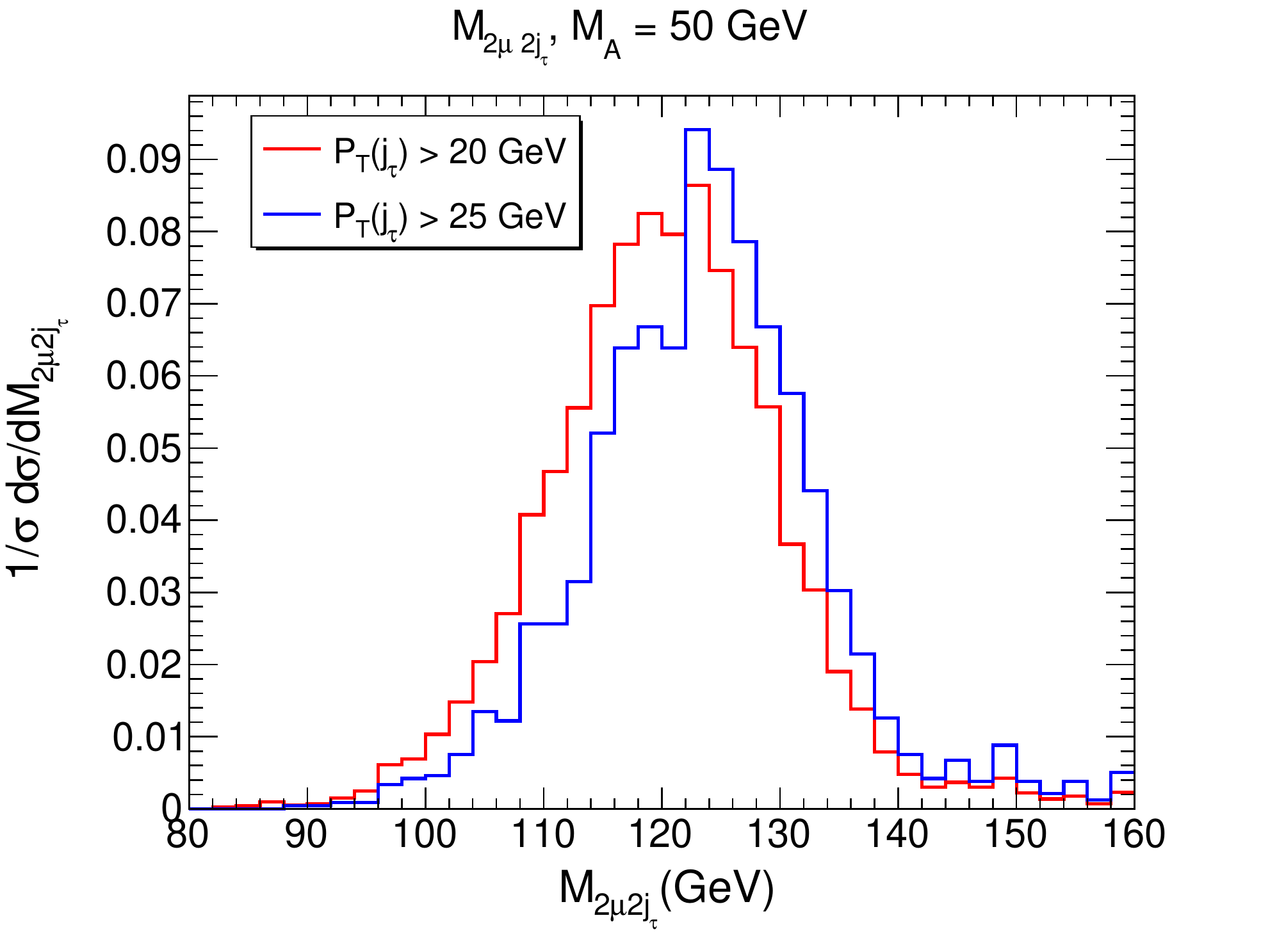}
   \includegraphics[width = 7.5cm]{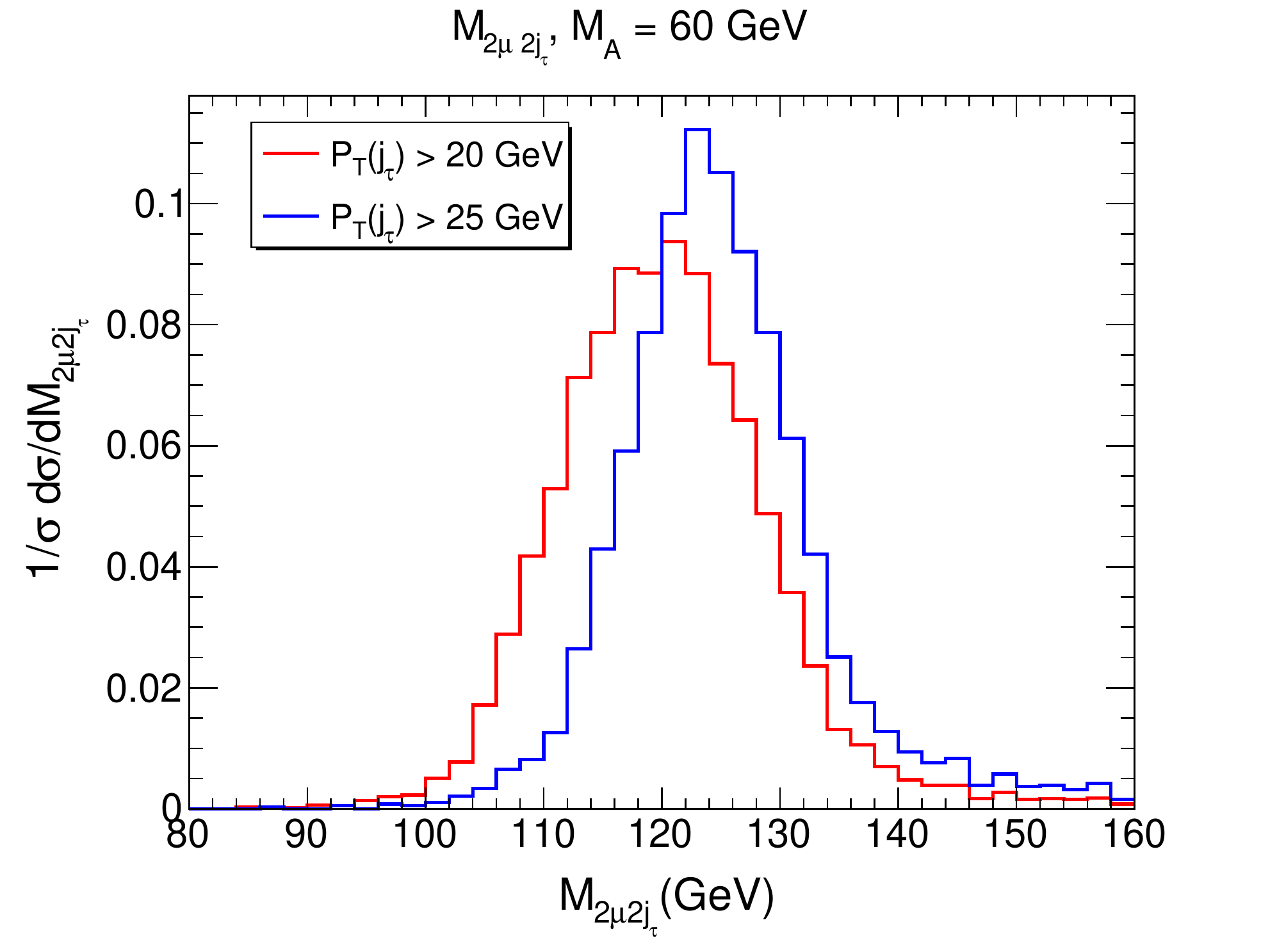}
   \caption{The invariant mass of the 2$\mu$ and 2 tau-tagged jets for $M_A = 50$ and $M_A = 60$.
   The figures illustrate how the higher $p_T (j_\tau)$ threshold leads to more
   precise reconstruction of the peak at $M_h = 125$ GeV.}
  \label{fig:IM_2mu2taujet}
  \end{figure}

  The hadronic decays of the $\tau$  are associated with some missing transverse energy  in the events. 
 For the signal events the $\tau$ leptons originate from the decay of a light 
 pseudoscalar ($A$) with mass 50 or 60 GeV. Hence, if the $p_T$ of the $\tau$-tagged jet has to be very 
 close to $m_A/2$, the corresponding missing energy in the final state is suppressed. The invariant 
 mass of the $\tau$-tagged jets will thus peak very close to the parent mass. 
 In Figure~\ref{fig:IM_taujets} we substantiate this claim by plotting  the invariant mass of  the
 $j_\tau\,j_\tau$ system for two different jet $p_T$ thresholds. One can clearly see that for  
 $p_T(j_\tau) > 25$ GeV the invariant mass peaks at the parent pseudoscalar mass, whereas 
 $M(j_\tau j_\tau)$ is peaking at a lower value than $M_A$ for  $p_T(j_\tau) > 20$ GeV. Also the 
 invariant mass peak is sharper for the higher $p_T$ threshold. The four-body invariant mass 
 $M_{2\mu 2j_\tau}$  also shows the same features and peaks close to $ M_h = 125 $ GeV as
 depicted in Figure.~\ref{fig:IM_2mu2taujet}. It is evident that these variables can be very
 efficient in minimizing the background events.

We use the following selection cuts to select our signal and reduce the accompanying backgrounds:
\begin{itemize}
 \item {\bf Preselection Cuts:} We require the final state to have two oppositely charged muons 
 of mininum $p_T > 10$ GeV and $|\eta| < 2.5$. We also require two tau-tagged jets ($j_\tau$) of minimum $p_T$, $p_T(j_\tau) > 20, 25$ GeV within $|\eta| < 2.5$. 
 \item The invariant mass of the di-muon system ($M_{\mu\mu}$) satisfies the window, $$| M_{\mu\mu}~-~ M_A | < 7.5  \textrm{ GeV}. $$
 \item The invariant mass of the two tau-tagged jets ($M_{j_\tau j_\tau}$) satisfies: 
 \begin{itemize}
 \item for $p_T(j_\tau) > 20$ GeV : $ (M_A - 20) < M_{j_\tau j_\tau} < (M_A + 10) $ GeV 
 \item for $p_T(j_\tau) > 25$ GeV : $|M_{j_\tau j_\tau} - M_A | < 15$ GeV.
 \end{itemize}
 \item The invariant mass of two muons and two taujets ($M_{2\mu 2j_\tau}$)lies within the range :
 \begin{itemize}
   \item for $p_T(j_\tau) > 20$ GeV : $ (M_h - 20) < M_{2\mu 2j_\tau} < (M_h + 10)$ GeV.
  \item for $p_T(j_\tau) > 20$ GeV : $|M_{2\mu 2j_\tau} - M_h| < 15$ GeV.
 \end{itemize}

 \end{itemize}
 
 Notice that we have taken asymmetric cut-windows with respect to $M_A$
 for $M_{j_\tau j_\tau}$ and $M_{2\mu 2j_\tau}$ for 
 $p_T(j_\tau) > 20$ GeV and symmetric ones for $p_T(j_\tau) > 25$ GeV. This has  to do with the fact that 
 for lower $p_T(j_\tau)$ cut, the invariant mass peaks at a lower value compared to the parent mass.

\section{Results and Discussion}
In Table~\ref{Table:Cut-flow-pt-20},  we present 
the cut flow for the signal and the various backgrounds  for the  benchmark points BP1 (BP2) where 
the number of events are calculated at the integrated luminosity of 3000 $ \rm{fb}^{-1}$. 
Note that some of the background events are estimated as upper bound (marked by an asterisk), 
as the number of simulated events passing the cuts drop down to very small values at 
some point in the cut flow table, even after simulating with $2\times 10^7$  events for
the background analysis. Since we adopt the Medium Tag point for tau-tagging, the mistagging rate
for a pair of light jets is $\sim ~10^{-4}$. This, along with a tight invariant mass window around $M_A$ helps 
to get rid of a major fraction of the various background channels. Demanding that 
$| M_{\mu\mu}~-~ M_A |<7.5$ GeV should take care of the $Z$ contribution in $p p \to \mu^+ \mu^- + jets $ and
$p p \to V V + jets $ . After the cut on $ M_{\mu\mu}$ only a feeble contribution from the photon 
(and partly off-shell $Z$) continuum can contribute in the  $p p \to \mu^+ \mu^- + jets $ channel.

\begin{table}[t]
\renewcommand{\arraystretch}{1.1}
\centering
\begin{tabular}{|c|c|c|c|c|}
\hline
% & & & & \\
Cuts                              & Signal    & $p p \to \mu^+ \mu^-$ & $p p \to V V  $ & $p p \to t \bar t $ \\ 
                                  &           & $+ jets$              & $+ jets $       & $+ jets$ \\
%                                   &           &                       &                 & \\
\hline\hline
\multicolumn{5}{|c|}{$p_T(j_\tau) > $ 20 GeV}\\
\hline\hline 
%                                   &           &                       &                 & \\
Preselection                      & 858 (1480) &  41041 (41041)	     &  107890 (107890) &  14486 (14486) \\ 
%                                   &           &                       &                 & \\
\hline
%                                   &           &                       &                 & \\
$| M_{\mu\mu}~-~ M_A |<7.5$ GeV & 836 (1430)  &  909 (779)           &  1189 (1325)     & 1637 (1697) \\ 
%                                   &           &                       &                 & \\
\hline
%                                   &           &                       &                 & \\
$M_{j_\tau j_\tau}>M_A - 20$ \& & 760 (1336) &    130 (390)          &  307 (654)       &  330 (419) \\ 
$M_{j_\tau j_\tau}<M_A+10$ GeV   &           &                       &                 & \\
%                                   &           &                       &                 & \\
\hline
%                                   &           &                       &                 & \\
$M_{2\mu 2j_\tau}>M_h-20$ \&       & 698 (1283) &  $<130$ ($<390)\ast$&  81 (109)         & 65 (51)  \\ 
$M_{2\mu 2j_\tau}<M_h+10$ GeV      &           &                       &                 & \\
%                                   &           &                       &                 & \\
\hline\hline
\multicolumn{5}{|c|}{$p_T(j_\tau) > $ 25 GeV}\\
\hline\hline
%                                   &           &                       &                 & \\
 Preselection                     &  277 (493) &  28833 (28833)      & 75209 (75209)   & 11629 (11629) \\ 
%                                   &           &                       &                 & \\
 \hline
%                                   &           &                       &                 & \\
$| M_{\mu\mu}~-~ M_A | < 7.5$ GeV &  269 (475) & 649   (390)         & 794 (924)       & 1324 (1396) \\ 
%                                   &           &                       &                 & \\
\hline
%                                   &           &                       &                 & \\
$| M(j_\tau j_\tau)-M_A |< 15$ GeV&  228 (420) & $<649$  (130)         & 112 (416)       & 182 (196) \\ 
%                                   &           &                       &                 & \\
\hline
%                                   &           &                       &                 & \\
$|M_{2\mu 2\,j_\tau} - M_h|<15$ GeV& 211 (410) &$<649\;(<130)\ast$    &  20 (15)         &  27 (27) \\ 
%                                   &           &                       &                 & \\
 \hline
\end{tabular}
\caption{Cut flow table for signal BP1(BP2) and different background processes with two different set of $p_T(j_\tau)$ 
  cuts as described in Section~\ref{sec:simulation}. The number of events are computed with integrated luminosity 
  of 3000 $\rm{fb}^{-1}$. The number of background events also depends on benchmark points as $M_A$ changes. }
\label{Table:Cut-flow-pt-20}
\end{table}

% \subsection{Discovery Potential}

\begin{figure}[t]
   \includegraphics[width = 7.5cm]{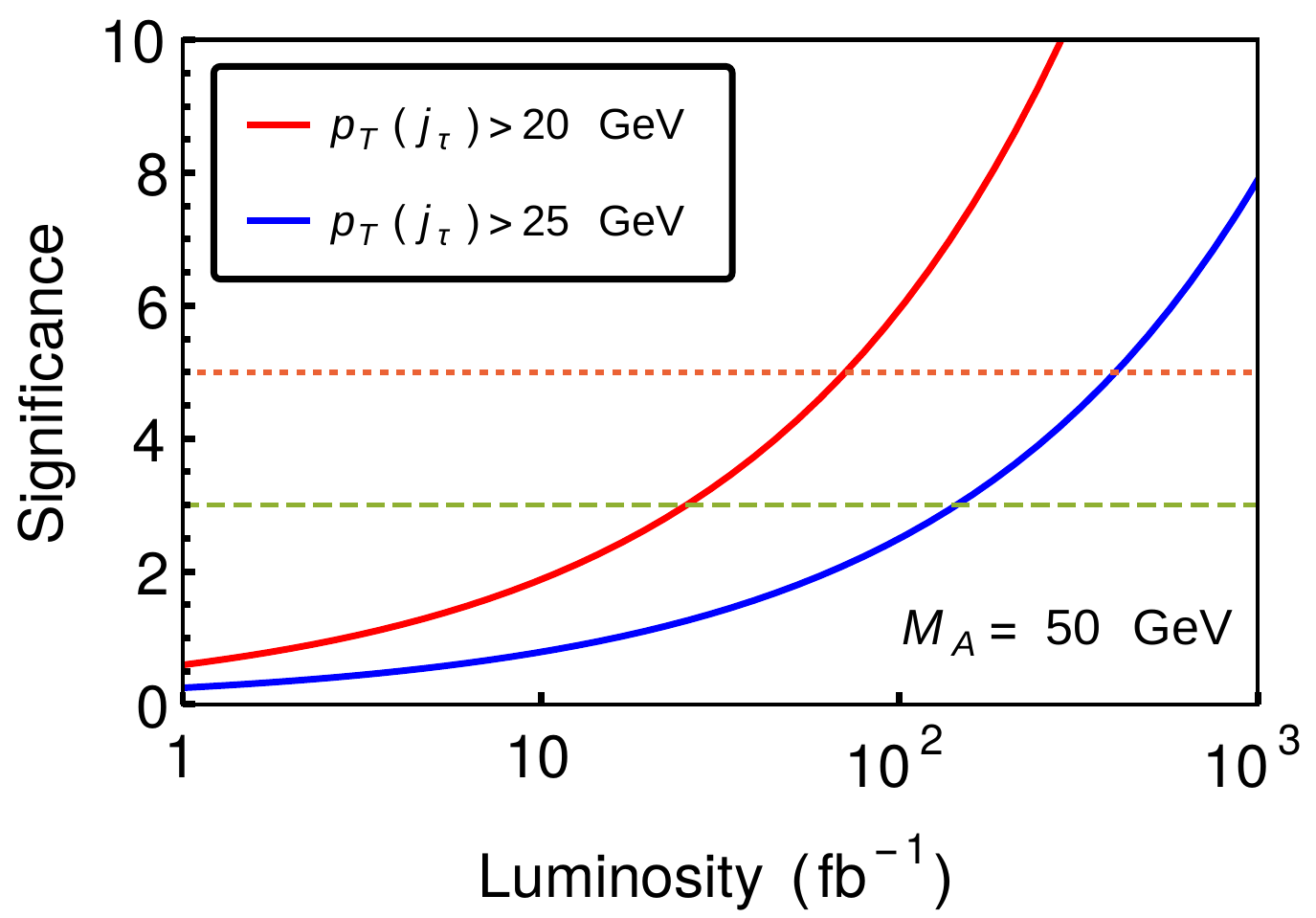}
   \includegraphics[width = 7.5cm]{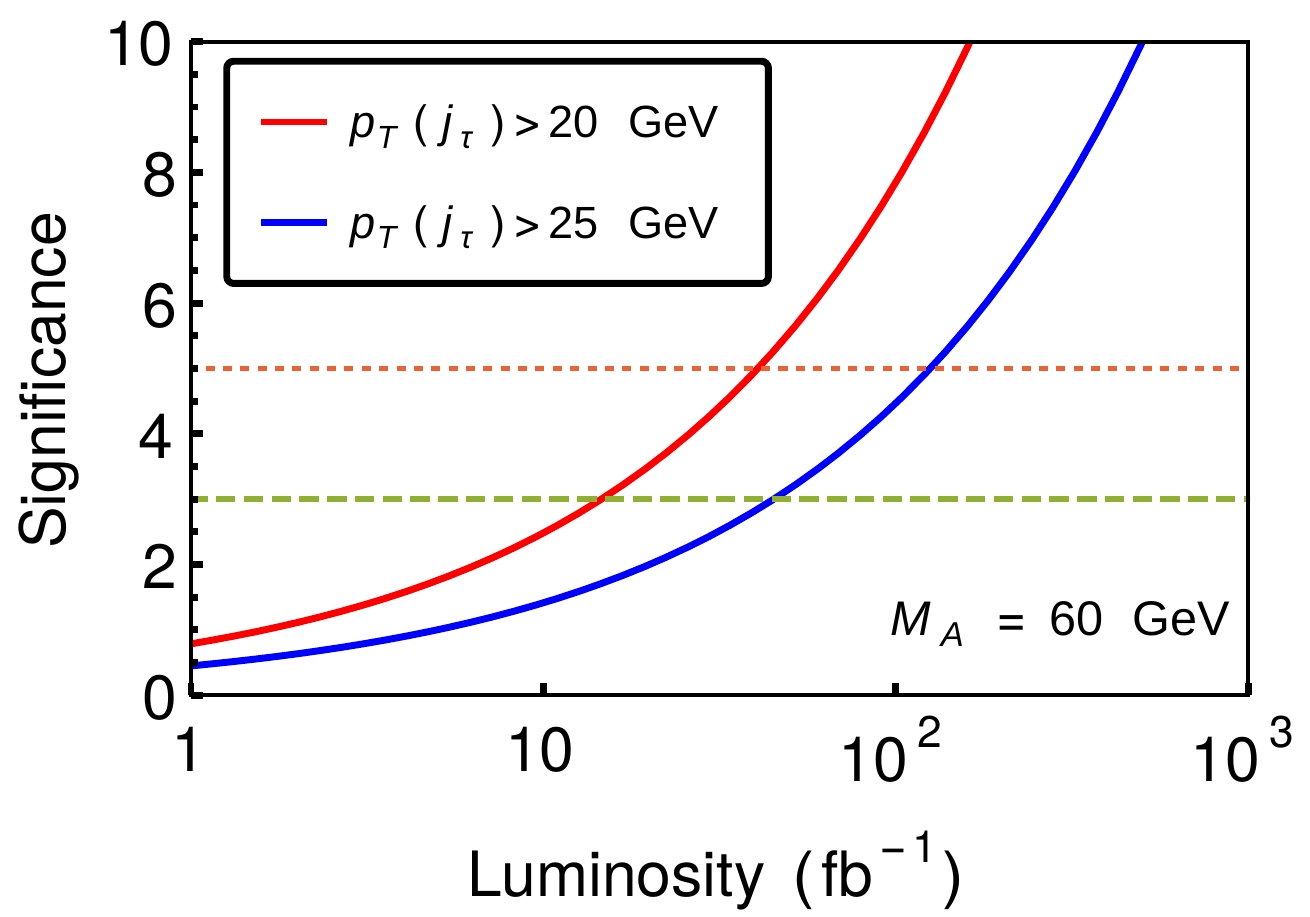}
   \caption{Discovery potential of the light pseudoscalar decaying to di-muon and di-tau channel 
   using invariant mass variables for BP1(left panel) and BP2(right panel) at 14 TeV LHC. }
  \label{fig:significance}
  \end{figure}

 We compute the statistical significance by using the formula,  
 $\mathcal{S} = \sqrt{2\left[(S+B)\textrm{ln}\left(1+\frac{S}{B}\right)-S\right]} $
 where $S(B)$ are number of signal (background) events which survive the cuts. 
 In Figure~\ref{fig:significance} we have plotted the significance $\mathcal{S}$ as a function 
 of integrated luminosity for both the benchmark points where BP1(BP2) corresponds to 
 $M_A = 50(60)$ GeV. For BP1 it is possible to reach 5$\sigma$ sensitivity 
 at integrated luminosity of 70(400) $\rm{fb}^{-1}$ with $p_T({j_\tau}) > 20 (25) $ GeV. 
  For BP2 the 5$\sigma$ sensitivity is achievable at 40(125) $\rm{fb}^{-1}$ integrated luminosity. 
Increasing the minimum $p_T(j_\tau)$ from 20 GeV to 25 GeV results in  better invariant mass 
peaks but provides fewer number of events which decreases the discovery prospect of the model. 
However the luminosity requirement is well within the reach of high luminosity run at the LHC.\\
 The benchmarks chosen in this work allowing for the $BR(h\to AA) \sim 15\%$ are 
close to the borderline of the exclusion limit on $\sigma(h) \times BR(h\to AA) \times {BR(A \to \mu\mu)}^2$ 
\cite{CMS-PAS-HIG-15-011, Aggleton:2016tdd}, when this is translated for $A \to 2\mu 2\tau$. 
However, one can still allow for a lower branching ratio for $h \to AA$, for example, close to 10$\%$, which keeps one well
within the exclusion limit, satisfying all the other constraints. This would entail the required luminosity
for a 5$\sigma$ discovery to be nearly double the values quoted above.

% \newpage
% \subsection{Determination of CP Properties}

 If $A$ were a scalar instead of a pseudoscalar, then it would also have decays into 
  $W^\ast W^\ast$ and $Z^\ast Z^\ast$  competing with the $\mu^+\mu^-$ mode. 
  The non-observation of such final states, even with accumulating luminosity,
  should act as a pointer to the $CP$-odd nature of $A$. 
  Secondly, the presence of such channels eats into the branching ratio of the A,
  and suppresses the $\mu^+\mu^-$ channel rate, reducing it below detectability. 
  The fact that we can reconstruct the $A$ via  the $\mu\mu$ peak (which is the main 
  point we make in this work) owes itself to the non-negligible branching ratio for 
  this mode, which would not have been possible if it were a scalar instead of a 
  pseudoscalar.

   On the other hand, if A were a superposition of a scalar and a pseudoscalar field
   (i.e. if CP were violated), then the taus coming from the decay of the other A
     would consist of unequal admixtures of right-and left-polarised states (both for
     $\tau^-$ and $\tau^+$). In principle, suitable triple products of vectors constructed
     out of the tau-decay products would have asymmetric distributions if CP-violation
     had taken place. However, the construction of such CP-asymmetric triple products
     would have required us to reconstruct the taus fully. This would warrant the so-called 
     collinear approximation, where the $\tau$, the decay product jet and 
     a neutrino would all move along the same straight line. This approximation is
     valid if the tau has an energy of at least about 40 GeV. In our case, for a
     light (50 -- 60 GeV) A this energy is not possessed by the taus, and thus
     their reconstruction is not reliable. Therefore, while one can distinguish a
     pure pseudoscalar from a pure scalar in this channel, identifying a
     CP-admixture is difficult.

It is  possible to search for the heavy scalar $H$ and the pseudoscalar using the
$pp\to Z \to HA$  production channel. In principle, this 
enables one to reconstruct the $H$ mass. However, this associated production rate  
will be two orders of magnitude smaller than the rate for $pp \to h \to AA$, principally
due to the large Higgs production rate from gluon fusion.  Nevertheless, if one notices a low-mass 
$\mu^+\mu^-$ peak from $A$, one can look for a tau-pair peak simultaneously in such events. It 
is relatively easy to reconstruct tau-leptons from the tau-jets in such a case, since these
taus are quite energetic and the collinear approximation~\cite{Rainwater:1998kj} will work
for them. Thus, in association with a light $A$ constructed in the way suggested
in our paper, a heavy $H$ can also be looked for, albeit at higher luminosity. 

%%%%%%%%%%%%%%%%%%%%%%%%%%%%%%%%%%%%%%%%%%%%%%%%%%%%%%%%%%%%%%%%%%%%%%%%%%%%%%%%%%%
 In addition, a light $A$ may of course be responsible for $4\tau$ final
states. Some channels leading to such a final state have been analyzed in \cite{Kanemura:2011kx,Kanemura:2014bqa,Chun:2015hsa}.
We observe that the $\geq 3\tau$ final state fares better in terms of the statistical 
significance owing to the dominant branching ratio of $A \to \tau^+ \tau^-$ as compared to
the much smaller branching ratio of $A \to \mu^+ \mu^-$.
For instance, for a 5$\sigma$ discovery of $M_A = 60$ GeV with $M_H =200$ GeV, 
the required luminosity is approximately 70 fb$^{-1}$ as against 218 fb$^{-1}$
for the $2\mu 2\tau$ final state.
 However, the di-muon pair is a lot cleaner to reconstruct, and gives an accurate
handle on the mass determination for the parent pseudoscalar. Thus the $2\mu 2\tau$ state 
is more informative when it comes to ``identifying'' the pseudoscalar.

%%%%%%%%%%%%%%%%%%%%%%%%%%%%%%%%%%%%%%%%%%%%%%%%%%%%%%%%%%%%%%%%%%%%%%%

\section{Summary and Conclusion}
While the Type-X 2HDM admits of a light pseudoscalar, the explicit reconstruction of its mass
is a challenging task. We propose to meet this challenge by making use of the small but 
non-negligible branching ratio for $A\to \mu^+\mu^-$, especially in the region of the 
parameter space, which best explains the muon anomalous magnetic moment. We have studied 
the channel $pp\to h\to AA\to\mu^+\mu^-\,\tau^+\tau^-$, with the taus decaying into 
a jet each. The $\mu^+\mu^-$ pair shows a conspicuous invariant mass peak at $M_A$. 
Besides, an appropriate $p_T$- cut on the tau-tagged jets also creates a $j_\tau j_\tau$ 
mass distribution that has a peak  in the neighbourhood of $M_A$. A proper window demanded 
of the latter invariant mass helps the effective tagging and background reduction 
for the $\mu^+\mu^-$ peak. We find that, for $M_A$ between 50 and 60 GeV, $M_A$ can 
be reconstructed in this manner, with statistical significance of 4-5 $\sigma$ 
with an integrated luminosity not far exceeding 100 $\rm{fb}^{-1}$ in the 14 TeV run.

\section{Acknowledgements}\label{sec:Acknowledgements}
BM thanks Korea Institute for Advanced Study for hospitality while this project 
was initiated. This work was partially supported by funding available from
the Department of Atomic Energy, Government of India, for the Regional Centre for
Accelerator-based Particle Physics (RECAPP), Harish-Chandra Research Institute. 
SD thanks N. Chakrabarty for helpful discussions.
 The authors acknowledge the use of the cluster computing setup available at 
 RECAPP and at the High Performance Computing facility of HRI.
% \newpage
%%%%%%%%%%%%%%
% \bibliographystyle{apsrev4-1}
% \bibliographystyle{unsrt}
\bibliographystyle{apsrev}
\bibliography{2hdm}
%%%%%%%%%%%%%%

\end{document}